\documentclass[11pt]{article}
\pdfoutput=1
\usepackage{ifpdf}
\usepackage{amssymb}
\usepackage{amsbsy}
\usepackage{graphicx}
\usepackage[numbers]{natbib}
\usepackage{color}
\usepackage{amsmath}

\title{Turbulence modulation in heavy-loaded suspensions of tiny particles}
\author{P.~Gualtieri,
\thanks{Email address for correspondence: paolo.gualtieri@uniroma1.it},
F.~Battista, \& C.M.~Casciola \\
Dipartimento di Ingegneria Meccanica e Aerospaziale\\
Sapienza Universit\`a di Roma}
\newcommand {\vxi}    {\mbox{\boldmath $\xi$}}
\newcommand {\vzeta}  {\mbox{\boldmath $\zeta$}}

\newcommand {\vD}     {{\bf D}}
\newcommand {\ve}     {{\bf e}}

\newcommand {\vF}     {{\bf F}}

\newcommand {\vk}     {{\bf k}}

\newcommand {\vvr}     {{\bf r}}

\newcommand {\vu}     {{\bf u}}
\newcommand {\vU}     {{\bf U}}
\newcommand {\vv}     {{\bf v}}

\newcommand {\vx}     {{\bf x}}

\newcommand {\vw}     {{\bf w}}

\def\Rset {{\rm I \kern-.2em R}} % blackboard bold R

\begin{document}
\maketitle

\begin{abstract}
The features of turbulence modulation produced by a heavy loaded suspension of 
small solid particles or liquid droplets are discussed by using a 
physically-based regularisation of particle-fluid interactions. The approach 
allows a robust description of the small scale properties of the system 
exploiting the convergence of the statistics with respect to the regularisation 
parameter. It is shown that sub-Kolmogorov particles/droplets  modify the 
energy spectrum leading to a scaling law, $E(k)\propto k^{-4}$, that emerges at 
small scales where the particle forcing balances the viscous dissipation. 
This regime is confirmed by Direct Numerical Simulation data of a 
particle-laden statistically steady homogeneous shear flow, demonstrating the 
ability of the regularised model to capture the relevant small-scale physics.  
The energy budget in spectral space, extended to account for the inter-phase 
momentum exchange, highlights how the particle provide an energy sink in the 
production range that turns into a source at small scales.  Overall, the 
dissipative fluid-particle interaction is found to  stall the energy cascade 
processes typical of Newtonian turbulent flows. In terms of particle 
statistics, clustering at small scale is depleted, with potential consequences 
for collision models. 
\end{abstract}

Particle laden turbulent flows are central in many physical and technological  
contexts. In astrophysics~\cite{Johansen_2007,Takeuchi_2002} the turbulence is 
known to influence the aggregation of dust particles in protoplanetary 
(accretion) disks, see~\cite{mitra2013can,pan2014turbulence,fu2014effects} and 
reference therein. 
Similarly, in warm clouds, the turbulence controls the growth by condensation 
of small droplets~\cite{lanotte2009cloud}, and ultimately speeds-up rain 
formation~\cite{Falkovich_2002,Shaw_2003}. In the combustion of liquid 
fuels~\cite{Marmottant_2004,lin_1998}, the turbulence determines the 
effectiveness of atomization, evaporation and mixing~\cite{jenny2012modeling}.
All these examples show that turbulence strongly interacts with the transported 
phase. Less understood is the reciprocal effect expected on the basis of the 
action-reaction principle by which the transported phase alters the turbulence. 
An extreme example of this reciprocal effect arises in the environmental 
context, where small active organisms such as plankton~\cite{Abraham_1998} 
or bacteria~\cite{dunkel2013fluid} induce small-scale chaotic flows 
which affects the chemical and the biological activity. Significant alteration of 
the turbulent flow is also found in bubbly grid-generated flows, 
\cite{lance1991turbulence}. In general, significant back reaction effects are 
expected in all the other contexts mentioned above. Concerning  in particular 
technological applications, in a typical diesel engine, 
see e.g.~\cite{ferguson2015internal}, the mass of fluid injected per cycle per 
cylinder in the form of small droplets is about $3\; \times 10^{-4} \, {\rm kg}$. 
Considering a four stroke, $2.5$ litre engine with $4$ cylinders, back of the 
envelope calculations immediately give a mass loading of about $\phi \simeq 0.4$ 
and a volume fraction of the order of $\phi_v \simeq 6\, \times  10^{-3}$. 
In modern common-rail injection systems the diameter of the droplets is about 
$d_p \simeq 0.1-10 \mu m$ whilst the Kolmogorov scale in a combustion chamber can be 
estimated on the order of  $\eta \simeq 30\mu m$. According to the accepted 
classification, see~ \cite{elghobashi1994predicting,elgo_map}, the suspension must 
then be considered dilute (no direct interaction among droplets) even though the 
inter-phase momentum coupling is particularly significant.

Among the different regimes of a particle laden flow~\cite{Balachandar_2010},
the present Communication addresses conditions like those mentioned above where 
i) the dimensions of the single 
suspended particle are much smaller than the relevant macroscopic scale of the 
turbulent flow, $d_p/\eta \ll1$; ii) the 
particles are extremely diluted with negligible direct particle-particle 
interaction, i.e. the volume fraction is small; iii) the mass loading of the 
suspension (particle to fluid mass ratio) is significant, 
$\phi = m_p/m_f = {\cal O}(1)$, implying that a considerable particle-induced 
force is exerted on the flow. In these conditions, beside turbulence-induced 
particle clustering already observed at small mass 
loading~\cite{bec2007heavy,saw2008inertial,young1997theory,fessler1994preferential,
chun2005clustering,Gualtieri_2009,picano2010anomalous}, new phenomena 
associated to turbulence modulation are expected, defining a still poorly 
understood realm of multiphase turbulence. 
In particular, the standard Kolmogorov-like paradigm~\cite{kolmogorov1941local}, 
which assumes that the turbulence is forced at large scales and eventually 
dissipated at small scales with a universal direct energy 
cascade~\cite{frisch1995turbulence}
emerging in the inertial range, is expected to fail.

In the new  conditions the particle population forces the fluid across the entire 
range of available scales, posing several new questions concerning the structure and 
the dynamics of turbulence under significant back-reaction effects. 
The first class of questions is methodological: how can the effect of many 
sub-Kolmogorov particles be modelled in a physically consistent manner in Direct 
Numerical Simulations (DNS)? Is a numerical simulation which truly couples the 
discrete, point-like phase with a continuum fluid feasible with the present 
state-of-the-art numerical tools? Can the coupling be made realistic yet 
affordable from the computational point of view? Are the singularities arising 
from the coupling amenable of rigorous treatment? As will be shown, answers 
to these methodological issues can be found in the context of a newly designed  
inter-phase momentum coupling strategy, the Exact Regularized Point Particle 
approach (ERPP)~\cite{Gualtieri_2015}. 
The second family of questions, is more physical: what are the effects of the 
back-reaction on the turbulence dynamics? How the disperse phase affects the 
energy  cascade processes and, in turns, the energy spectrum? What is the 
resulting effect of the coupling on the particle population?  
Can we trust the numerical predictions, particularly at small scales, where 
most of the particle-fluid interaction is expected to occur?

This Communication provides an answer to all these questions, discussing the 
results of new simulations based on the ERPP approach that are free of the bias 
that hampers other available techniques aimed at realising the particle-fluid 
interaction. Among others, the crucial advancements over the present 
state-of-the-art concern: a) a physically-based, grid-independent 
regularisation of the singular response of point-like particles; b) the 
possibility to take a weak limit for the statistics with the regularisation 
parameter approaching zero; c) the ability to exactly remove from the field 
the unphysical self-induction velocity of each single particle in the 
calculation of the hydrodynamic force; d) the recovery of the exact momentum 
balance in the force coupling of each particle with the fluid; e) the 
convergence of the coupling scheme also when a fixed  number of particles, 
independent of grid size, is considered.

In order to address these issues in the cleanest form, the flow should be as simple 
as possible. Traditionally homogeneous and isotropic turbulence is the elective 
choice. However it requires an external forcing acting at large scales to provide 
the energy dissipated by viscosity. Although this is not an issue for classical 
Newtonian turbulence, the external forcing introduces undesired features in the 
context of particle laden flows in presence of backreaction. The reason is that, as 
shown below, the particle forms long clusters spanning the entire range of scales, 
up to the integral scale. The external forcing interferes with the large scales of 
the clusters and their backreaction on the fluid, thereby introducing dynamical 
artefacts. A flow able to self-sustain the turbulence with no artificial external 
forcing which still retains a substantial simplicity, e.g. statistical spatial 
homogeneity and stationarity, is the homogeneous shear flow, where a linear average 
shear is enforced on turbulence fluctuations. This flow, described in detail in 
the Supplemental Material (SM, \cite{SI}), will be exploited below to discuss 
generic features of particles laden flows under strong loading.

When $d_p \ll \eta$, the carrier flow is described by the incompressible 
Navier-Stokes equations 
%----------------------------------------------------------------------------------
%----------------------------------------------------------------------------------
\begin{equation}
\label{eq:NS}
\begin{array}{l}
\displaystyle \nabla \cdot {\bf u} = 0 \\ \\
\displaystyle \rho \left( \frac{\partial {\bf u}}{\partial t} + {\bf u} \cdot 
\nabla {\bf u} \right) = -\nabla p  + \mu \nabla^2 {\bf u} + {\bf F}
\end{array}
\end{equation}
%----------------------------------------------------------------------------------
where 
%----------------------------------------------------------------------------------
\begin{equation}
\label{eq:F_singular}
{\bf F}({\bf x},t)=
-\sum_{p=1}^{N_p} {\bf D}_p(t) \delta\left[{\bf x} - {\bf x}_p(t) \right]
\end{equation}
%----------------------------------------------------------------------------------
is the (singular) field representing the back-reaction of the point-like particles
on the flow. In equation (\ref{eq:F_singular}), $N_p$ denotes the number of 
particles, ${\bf D}_p$ the hydrodynamic force acting on the p-{\rm th} particle 
and the Dirac delta function localises the force at the particle position 
${\bf x}_p(t)$. Clearly equations (\ref{eq:NS}-\ref{eq:F_singular}) need to be 
regularized to be amenable to numerical treatment. 

In the classical Particle in Cell approach, see e.g. \cite{crowe1977particle}, the 
singularity is removed averaging the feedback on the computational cell, giving rice 
to several drawbacks, 
see e.g.~\cite{Gualtieri_2013,Balachandar_2010,boivin1998direct}.
Typically, convergence can be achieved only at constant number of particles per 
computational cell, implying that the number of particles should increase 
(at constant mass loading) as the grid size is reduced. Additionally, the particles 
are affected by their own self-induced disturbance, which introduces errors in 
the hydrodynamics force. This source of error  gets more and more pronounced as the
number of particles per cell is reduced, as always happens under grid refinement.
These drawbacks do not affect the ERPP method where the Dirac delta function is 
regularized in a physically consistent manner. The disturbance due to each 
point-like particle is evaluated in a closed analytical form exploiting the exact 
solution of a local unsteady Stokes problem and the viscosity of the fluid naturally 
takes care of regularising the fluid response to the particle forcing.
In turbulence, when $d_p \ll \eta$, is natural to set the regularisation length 
on the order of the Kolmogorov length-scales $\eta$ or below.  The singular 
forcing (\ref{eq:F_singular}) is effectively replaced by its (exact) regularized 
counterpart, 
%----------------------------------------------------------------------------------
%----------------------------------------------------------------------------------
\begin{equation}
\label{eq:F_regular}
{\bf F}_R({\bf x},t)=-\sum_{p=1}^{N_p} {\bf D}_p(t-\varepsilon_R) 
g\left[{\bf x} - {\bf x}_p(t-\varepsilon_R),\varepsilon_R \right] \, ,
\end{equation}
%----------------------------------------------------------------------------------
%----------------------------------------------------------------------------------
where the Gaussian function $g$ consistently emerges from the small scale diffusion 
of the particle disturbance field described by the unsteady Stokes 
operator~\cite{Gualtieri_2015}.
The spatial cut-off scale $\sigma_R=\sqrt{2\nu\varepsilon_R}$ is directly 
related to the diffusion time-scale $\varepsilon_R$ which represent the typical 
time needed by the singular vorticity produced by the particle at time 
$t-\varepsilon_R$ to spread over the resolved length scale at time $t$, 
see~\cite{Gualtieri_2015} and Supplemental Material~\cite{SI}.

The dispersed particles follow Newton's equations,
%----------------------------------------------------------------------------------
%----------------------------------------------------------------------------------
\begin{equation}
\label{eq:Newton}
\begin{array}{l}
\displaystyle \frac{d {\bf x}_p }{d\,t }={\bf v}_p\\ \\
\displaystyle \frac{d {\bf v}_p }{d\,t }=\frac{{\bf D}_p}{m_p} = 
\frac{{\bf u}\big|_{{\bf x}_p} - {\bf v}_p}{\tau_p}\, ,
\end{array}
\end{equation}
%----------------------------------------------------------------------------------
%----------------------------------------------------------------------------------
%---------------------------------------------------------------------------------
%---------------------------------------------------------------------------------
\begin{figure}[b!]
\centering
\includegraphics[width=0.6\textwidth]{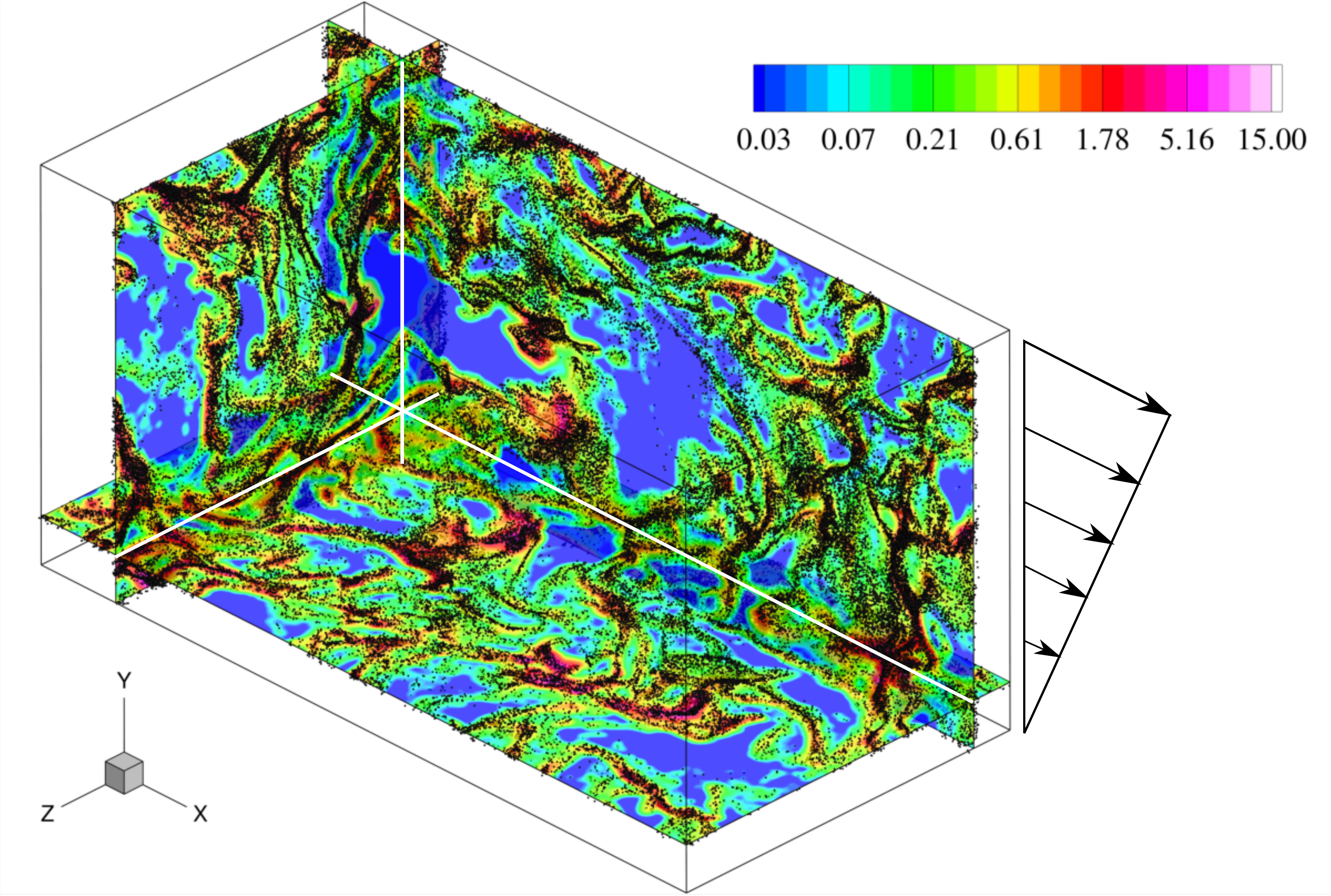}
\caption{\label{fig:inst} Snapshot of the instantaneous particle configuration 
(scatter plot) and of the force feedback field, $\|{\bf F}_R\|$
exerted by the particles on the fluid (contour plot). The slice in the  $x-y$ plane
is few Kolmogorov-lengths thick. The mean flow $U(y) = S\, y$ is from left to 
right.
}
\end{figure}
%---------------------------------------------------------------------------------
%---------------------------------------------------------------------------------
where ${\bf x}_p$ and ${\bf v}_p$, $p=1, \dots, N_p$, are the particle positions 
and velocities, respectively, $m_p$ is the particle mass and, in the conditions 
considered here, ${\bf D}_p$ reduces to the Stokes 
drag~\cite{gatignol1983faxen,Maxey_1983} proportional to 
the fluid-particle relative velocity with ${\bf u}\big|_{{\bf x}_p}$ the fluid 
velocity at the particle position. In the jargon of particle laden flows, 
such relative velocity is
sometimes called the slip velocity.  Since the particle modifies the fluid velocity, 
care should be taken not to contaminate ${\bf u}\big|_{{\bf x}_p}$ with the 
particle self-disturbance. Otherwise, at decreasing the grid size, the spurious 
contribution would dominate the overall particle-fluid interaction.
The Stokes number, $St_\eta=\tau_p/\tau_\eta$,  where $\tau_\eta$ 
is the Kolmogorov time scale of the turbulence and 
$\tau_p=(\rho_p/\rho) d_p^2/(18 \nu)$ is the Stokes relaxation time, is a central 
control parameter which, e.g., determines the intensity of particle clustering, 
that is the trend to segregate~\cite{bec2007heavy,chun2005clustering,Falkovich_2002,Gualtieri_2009,Gualtieri_2013,lanotte2009cloud,saw2008inertial}
in long, tiny structures. 

Figure~\ref{fig:inst} shows a slice of an instantaneous configuration of particle 
distribution and feedback force field in a turbulent homogeneous shear flow. 
The turbulence, at $Re_\lambda= \lambda u_{{\rm rms}}  /\nu= 80$, 
with $\lambda = u_{{\rm rms}}\sqrt{\nu/\epsilon}$ and
$u_{{\rm rms}} = \sqrt{ \langle (u-{\bar u})^2 \rangle}$,  
$St_\eta=1$ and $\phi=0.4$, is sustained by a constant mean shear $S = dU_x/dy$,
see Supplemental Material~\cite{SI} for details. The energy is extracted from the mean 
flow by the Reynolds 
shear stresses $-\langle u \, v \rangle$ which force the turbulent fluctuations at 
scales larger than the shear scale $L_S= \sqrt{\epsilon/S^3}$ \cite{casciola2003scale}.
Typical of unitary Stokes number flows, the disperse phase forms elongated
clusters, apparent in the plot. They are oriented by the mean flow 
which imprints on them a strong anisotropy. The clusters span a range of scales from 
their width, of the order of the dissipative scale, up to their length, comparable 
with the integral scale of the flow~\cite{Gualtieri_2009,Gualtieri_2013}. 
The force feedback ${\bf F}_R$ is strongly correlated with the clusters and affect 
the same range of scales. This kind of distributed, effective field differs 
substantially from the classical Kolmogorov scenario where the forcing is designed  
to prevent the flow from dissipating, it is confined to the large scales to avoid 
contamination of the cascade and is assumed to be statistically independent of 
the flow.

It is instrumental to look at the flow in spectral space where,
adopting index notation, the interphase momentum coupling is 
described by the Fourier transform $\cal F$ of the correlation
$\Psi_{ij}(\vk)={\cal F}\langle F_{R,i}(\vx) \, u_j(\vx+\vvr) \rangle$
between the back-reaction and the fluid velocity. The quantity
$\Psi(k) = \int_\Omega \Psi_{ii}(\vk) \, k^2 d\Omega$,
where the integral is taken over the solid angle $\Omega$ in wavenumber space and 
$k^2= \vk \cdot \vk$, forces the equation for the turbulence spectrum $E(k)$ 
according to 
%---------------------------------------------------------------------------------
\begin{equation}
\label{eq:k_budget}
\frac{\partial E(k)}{\partial t}= T(k) + P(k) - D(k) + \Psi(k)  \,
\end{equation}
%---------------------------------------------------------------------------------
where  $E(k) = \int_\Omega E_{ii}(\vk)  \, k^2 d\Omega$ and
$E_{ij}(\vk)={\cal F}\langle u_i(\vx) \, u_j(\vx+\vvr) \rangle$.
Equation~\eqref{eq:k_budget} is the extension to particle laden flows of the 
classical equation for the spectral balance of turbulent kinetic energy, sometimes 
called the Kolmogorov-Onsager-von Weizs\"acker-Heisenberg 
equation~\cite{eyink2006onsager,frisch1995turbulence}.
In equation (\ref{eq:k_budget}) the energy transfer term $T(k)$ is defined 
as $T(k)=\int_\Omega \imath k_j T_j(\vk) \, k^2 d\Omega$ where the 
Fourier transform of the triple correlation function is
$T_j(\vk)={\cal F}\langle u_i(\vx) u_i(\vx+\vvr) u_j(\vx) -
u_i(\vx+\vvr) u_i(\vx) u_j(\vx+\vvr) \rangle$.
The non-linear triadic interactions among different Fourier modes conserves energy,
$\int_0^\infty T(k) \, dk =0$, and ultimately originate the energy cascade. 
$P(k)=-S E_{uv}(k)$ is the production of turbulent kinetic energy at wavenumber 
$k$ where $E_{uv} = E_{12}(k)$ is the energy cospectrum and $D(k)=2 \nu k^2 E(k)$ 
the dissipation spectrum. Note that once integrated across the entire range of 
wavenumbers the energy cospectrum returns the Reynolds shear stresses
$-\langle u \, v \rangle=\int_0^\infty E_{uv}(k) \, dk$, and the
dissipation spectrum gives the viscous dissipation
$\epsilon = \int_0^\infty D(k) \, dk$. In statistically steady conditions
the time derivative of the energy spectrum vanishes. 

%---------------------------------------------------------------------------------
%---------------------------------------------------------------------------------
\begin{figure}[t!]
\includegraphics[width=.49\textwidth]{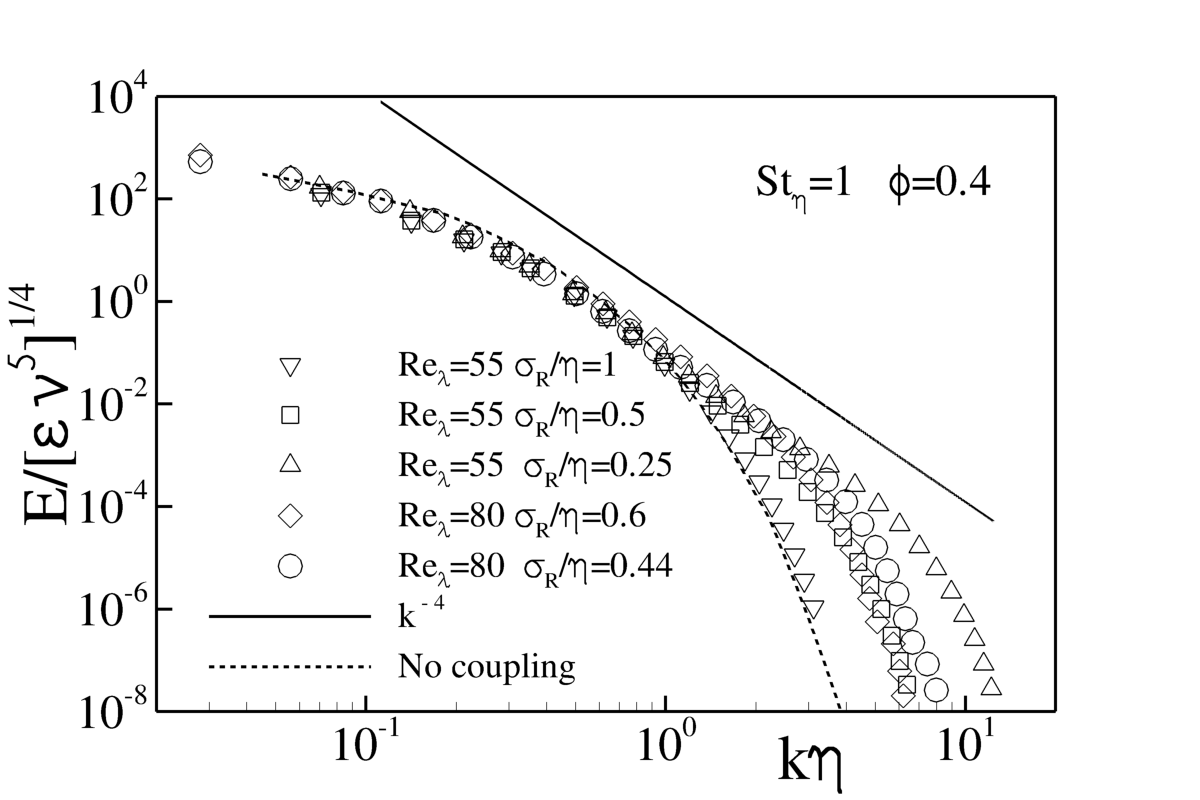}
\includegraphics[width=.49\textwidth]{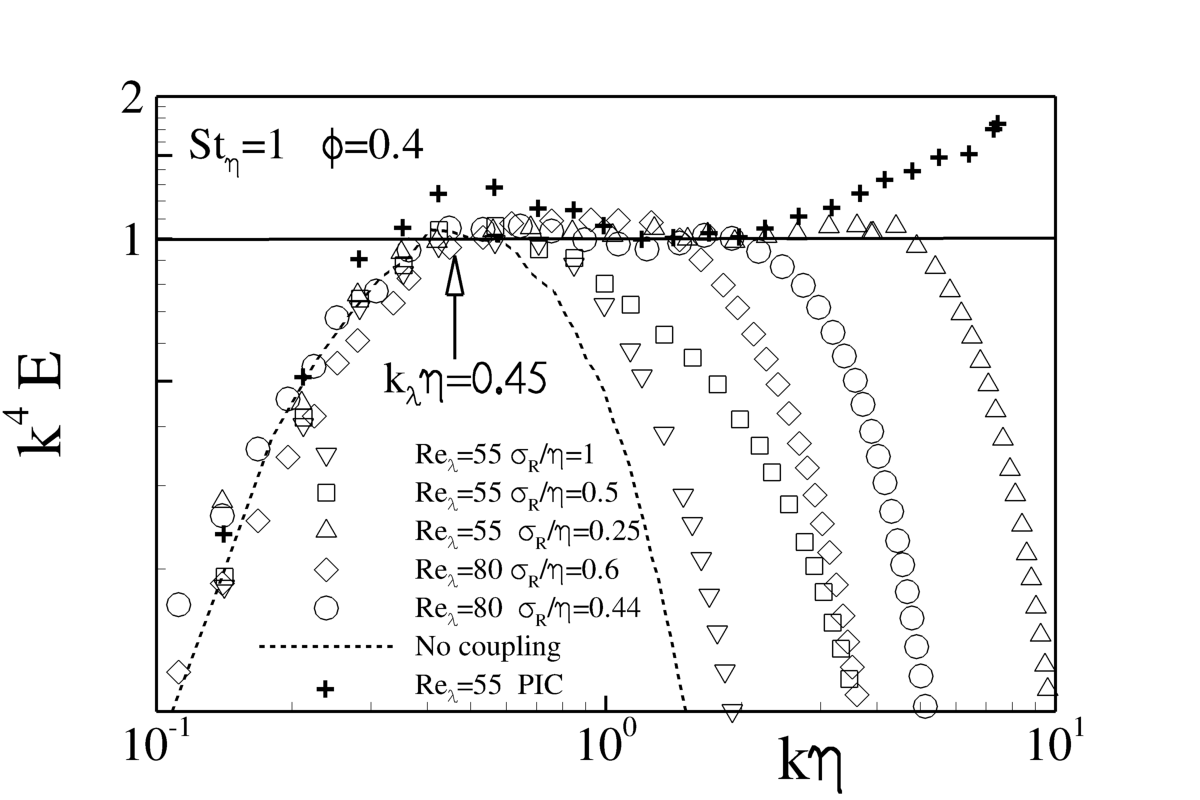}
\caption{Left panel: energy spectra $E(k)$ in Kolmogorov unitis versus normalised 
wave number $k\eta$.
Right panel: compensated energy spectra $k^4 \, E(k)$ v.s. $k\eta$, here
$E(k)$ is in arbitrary units to collapse the scaling plateau.
Data at $Re_\lambda=55$: $\sigma_R/\eta=1$ ($\bigtriangledown$); 
$\sigma_R/\eta=0.5$  ($\square$); $\sigma_R/\eta=0.25$ ($\triangle$).
For the three cases the resolution of the DNS is $192 \times 96 \times 96$;
$384 \times 192 \times 192$ and $768 \times 384 \times 384$ Fourier modes.
Data at $Re_\lambda=80$: $\sigma_R/\eta=0.6$ ($\diamond$);
$\sigma_R/\eta=0.4$ ($\bigcirc$). For the two cases DNS resolution is 
$768 \times 384 \times 384$ and $1024 \times 512 \times 512$ Fourier modes 
respectively. In all cases the computational box is $4\pi \times 2\pi \times 2\pi$
with a regularisation length-scale $\sigma_R=\Delta$ where $\Delta$ is the 
grid spacing in physical space. The solid line corresponds to the scaling law 
$E(k)\propto k^{-4}$ and the dashed lines reports data for the uncoupled case 
(no back-reaction on the fluid). In the right panel data at
$Re_\lambda=55$ obtained with the PIC approach (${\bf +}$ symbols) have been 
reported for comparison.
\label{fig:spectra}
} 
\end{figure}
%---------------------------------------------------------------------------------
%---------------------------------------------------------------------------------
Concerning eq.~\eqref{eq:k_budget},
one of the simulative issues with particle laden flows in the two coupling regime,
is the sensitivity of small scale observables to the numerical implementation of the 
particle feedback.  The approach here proposed allows for obtaining a clean 
asymptotic also for small scale observables.
This is achieved in the limit $\sigma_R \to 0$, where the limit is to be understood 
in the weak sense, i.e. first the statistics is acquired as a function of the 
regularisation parameter and only after the limit is taken on the averages.
This process is illustrated in figure~\ref{fig:spectra} where turbulent kinetic 
energy spectra are shown for the same particle population and two different 
Reynolds number at decreasing $\sigma_R/\eta$. Apparently the data nicely 
collapse and a well defined energy distribution emerges at decreasing $\sigma_R$. 
This is expected at large scales which soon become independent of the 
regularisation parameter. A new feature emerges at small scales (large wavenumber)
where a well definite scaling range eventually appears at $k \eta \simeq 1$.
The right panel shows the compensated plot, $k^4 E(k)$ vs 
$k\eta$. About one decade of $k^{-4}$ scaling is detected for the 
smallest $\sigma_R/\eta$ we have considered. The scaling range approximately 
extend from about $k_\lambda \eta \simeq 0.45$, which is order of the Taylor 
micro-scale where the dissipation spectrum peaks, to the 
cut-off $k \sigma_R \simeq 1$ corresponding to $k\eta \simeq 4$. 
%---------------------------------------------------------------------------------
\begin{figure}[b!]
\includegraphics[width=.49\textwidth]{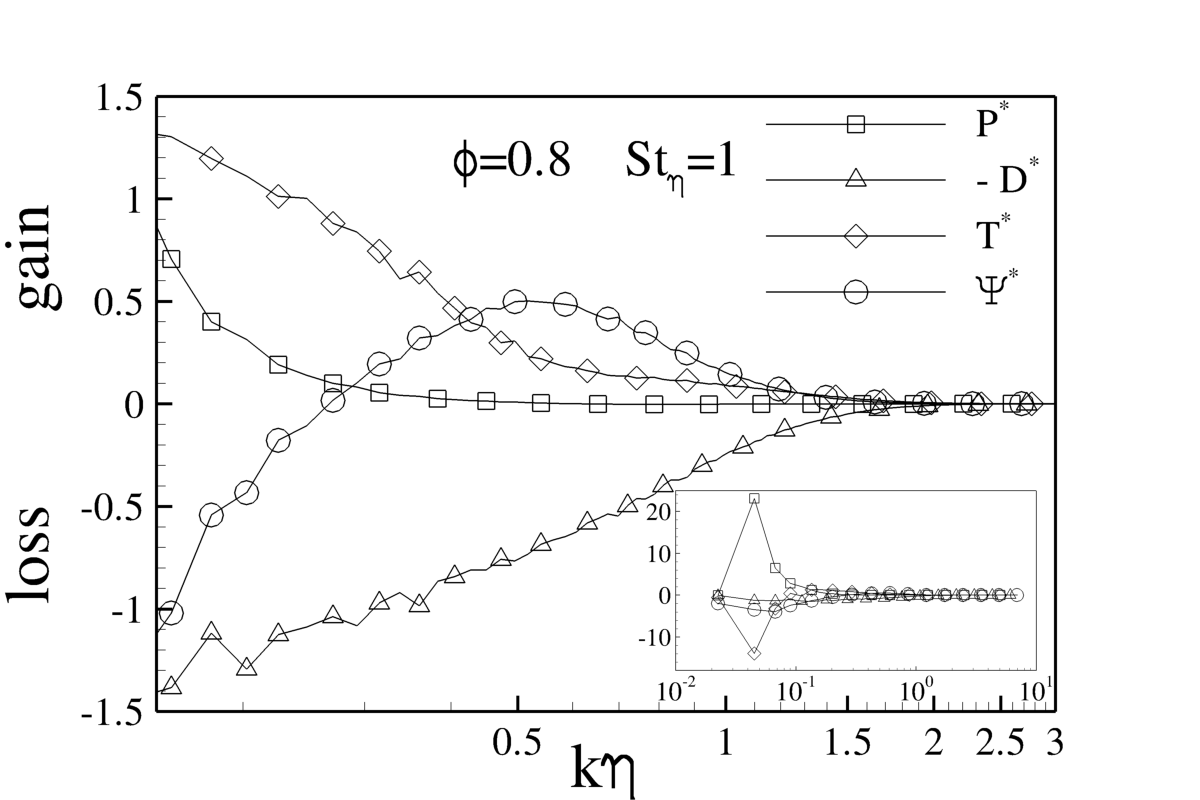}
\includegraphics[width=.49\textwidth]{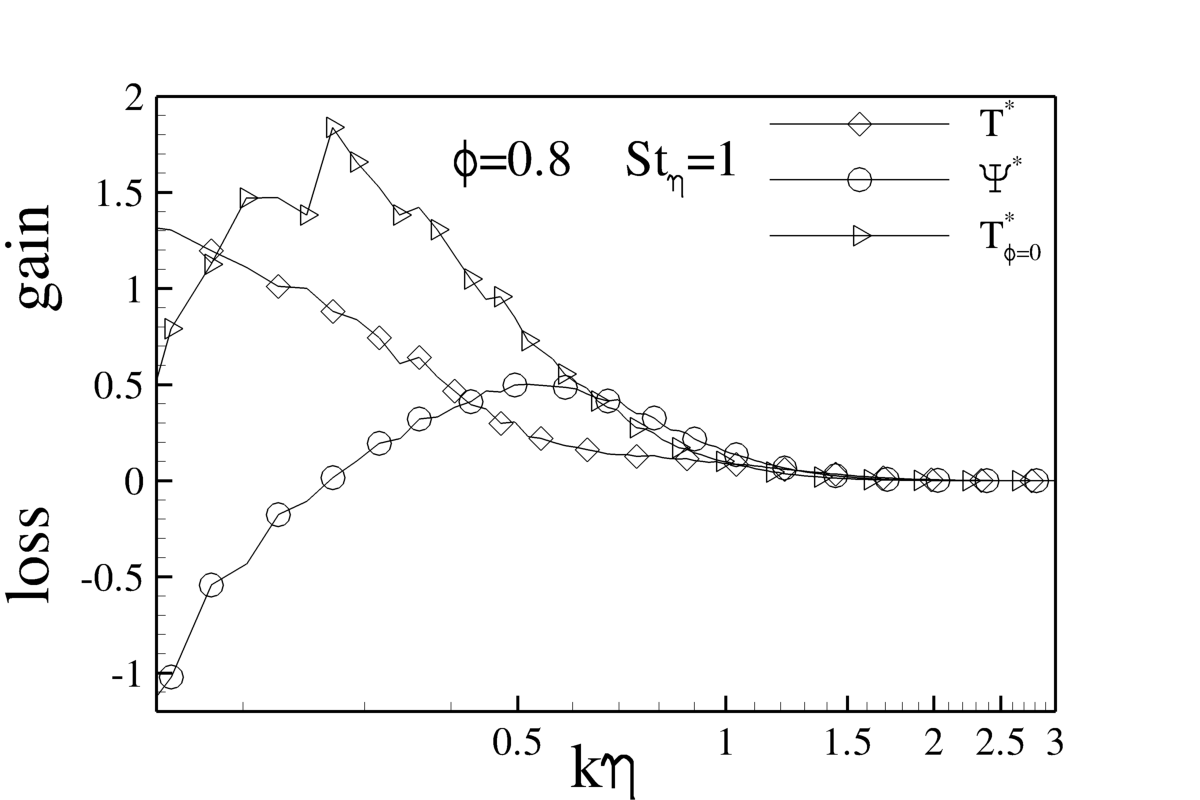}
\caption{Left panel: scale-by-scale energy budget (\ref{eq:k_budget}) in 
spectral space for the case at $Re_\lambda=80$, $St_\eta=1$ and $\phi=0.8$. 
Transfer $T(k)$, ($\diamond$); production $P(k)$, ($\square$); dissipation $D(k)$, 
($\triangle$); inter-phase coupling $\Psi(k)$, ($\bigcirc$). 
Main panel: close up view of the range of scales where the scaling law 
$E(k)\propto k^{-4}$ is measured, see figure \ref{fig:spectra}. 
Inset: representation of the budget in the whole range of scales.
Right panel: the transfer term $T(k)_{\phi=0}$ in the uncoupled case ($\rhd$) 
is compared against $T(k)$ in the coupled case. 
The asterisk denotes normalisation with respect Kolmogorov units, i.e.
$T^*=T/\left(\nu \epsilon \right)^{3/4}$, $P^*=P/\left(\nu \epsilon \right)^{3/4}$,
$D^*=T/\left(\nu \epsilon \right)^{3/4}$, $\Psi^*=\Psi/\left(\nu \epsilon \right)^{3/4}$.
\label{fig:spectral_budget}
} 
\end{figure}
%---------------------------------------------------------------------------------
We may note that data in absence of particle 
feedback show a completely different trend, consistently with the 
behaviour expected in the dissipation range. This result shows that the regularisation 
procedure we have put forward can be used to obtain physically significant and 
numerically convergent information on the small scale statistics 
of the system. Indeed, by reducing $\sigma_R$ at given turbulence intensity, 
we can approach any given small scale in the system.
This is important in view of taking into account interactions between particles, 
such as collisions, lubrication effects, short range attraction or repulsion between  
particles, e.g. Van der Walls forces, which arise at the inner length scale $d_p$ of 
the particles.

For comparison, the right panel of figure~\ref{fig:spectra} reports the 
compensated spectra obtained with the PIC approach operated in the same conditions, 
namely $Re_\lambda=55$ and  $\sigma_R/\eta=0.5$. Mass loading $\phi=0.4$ and Stokes 
number  $St_\eta=1$ fix the number of particles $N_p=595520$,  corresponding to  
few particles per cell, namely $N_p/N_c \simeq 0.04$ where $N_c$ is the number of 
computational cells. The PIC approach is reasonably able to describe the behaviour of 
the compensated spectrum at $k\eta \simeq 1$ where a glimpse of a short plateau 
seems to appear. However, at  smaller scales, the trend reveals a clear departure 
from the $k^{-4}$ scaling law. The reason is that the high wave number modes are 
badly behaved due to the non-smooth and grid dependent numerical feedback field, 
see e.g.~\cite{Gualtieri_2013}. This hampers reaching progressively smaller and 
smaller scales. The behaviour gets worser and worser when finer grids are used 
(data not shown).

The spectral budget, eq.~(\ref{eq:k_budget}), is shown in 
figure~\ref{fig:spectral_budget}. The main panel focuses on the range of  
wave-numbers where the $k^{-4}$-scaling is observed (see the inset for a global view).
The production $P(k)$ and the transfer term $T(k)$ vanish where $k\eta \simeq 1$, 
showing that the dominant balance is between the inter-phase coupling 
$\Psi(k)$ -- the only energy source present at those scales in absence of the 
energy transfer -- and the viscous dissipation $D(k)$. The back-reaction has 
overwhelmed the inertial transfer and stalled the energy cascade, right panel with 
the comparison of the energy transfer with, $T(k)$, and without, $T(k)_{\phi=0}$, 
coupling. The reduced transfer is replaced by the energy injected by the 
particles which, in turn, drain from the large scales the energy $P(k)$ extracts from 
the mean flow. As a consequence, the energy feeding the cascade is reduced by the 
amount drained by the disperse phase. The balance between energy intercepted by the 
particles at large scales and the energy released at small scales is negative,
%---------------------------------------------------------------------------------
\begin{equation}
\nonumber
\int_0^\infty \Psi(k) \, dk = -\epsilon_e < 0
\end{equation}
%---------------------------------------------------------------------------------
implying a dissipative effect of the particles.
Considering the overall budget, including fluid and particles, 
$-S\langle u \, v \rangle= \epsilon + \epsilon_e$, the energy produced by the 
Reynolds stresses is turned into the sum of 
viscous dissipation and the extra-dissipation due to the particles,
$\epsilon_e$. In other words, the disperse phase provides an alternative 
dissipation channel.
%---------------------------------------------------------------------------------
\begin{figure}[t!]
\centering
\includegraphics[width=.49\textwidth]{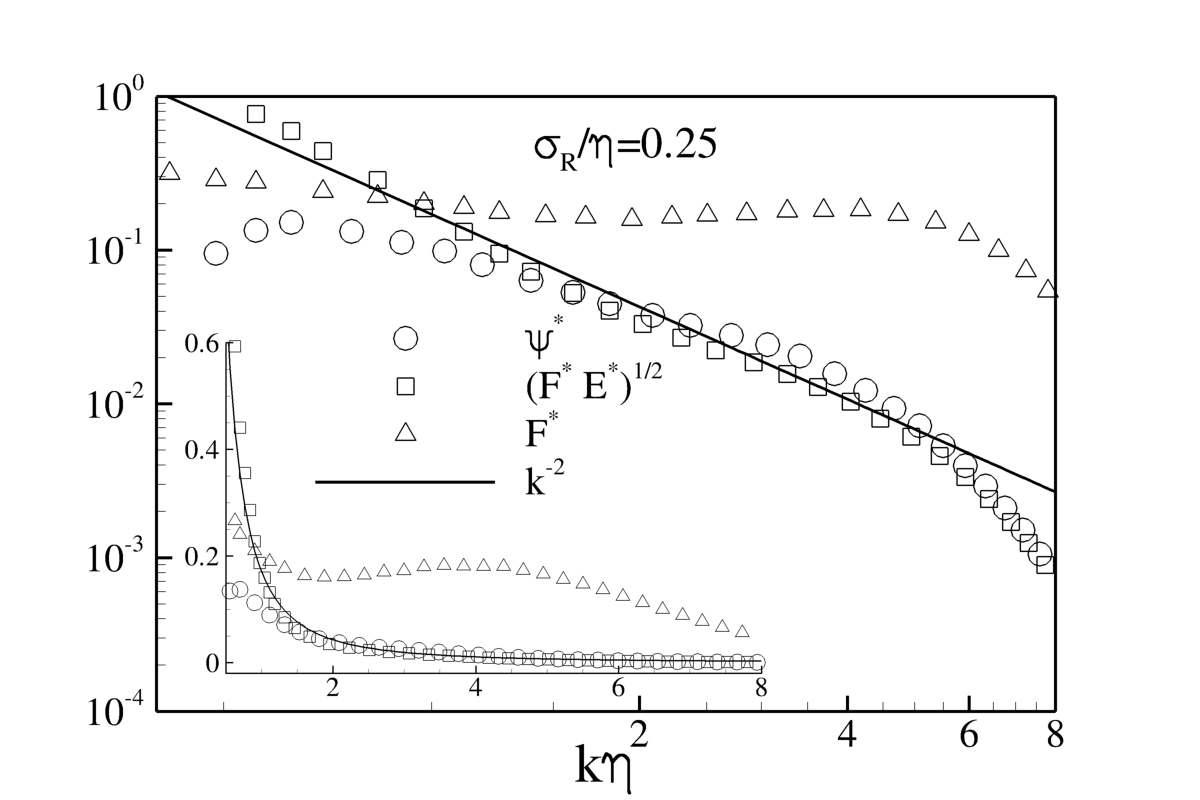}
\caption{Data at $Re_\lambda=55$ for $\sigma_R/\eta=0.25$.
Inter-phase coupling $\Psi(k)$, ($\bigcirc$), spectrum of the particle back-reaction
field $F(k)$ ($\triangle$) and $\sqrt{F(k) \, E(k)}$ ($\square$) in spectral space. 
The solid line denote the $k^{-2}$ scaling law. Inset: same data of 
the main panel in a lin-lin plot.
The asterisk denotes normalisation with respect Kolmogorov units, i.e.
$\Psi^*=\Psi/\left(\nu \epsilon \right)^{3/4}$,
$F^*=F/\left(\nu \epsilon \right)^{3/4}$, $E^*=E/\left(\nu \epsilon \right)^{3/4}$,
\label{fig:spectral_corr}
The spectrum of the particle back-reaction field $F(k)$ ($\triangle$) is in arbitrary 
units to be compared with the other terms in the budget.
} 
\end{figure}
%---------------------------------------------------------------------------------

The data just discussed show that the $k^{-4}$ scaling range corresponds to the 
region where $\Psi(k) \simeq D(k)$. Note that in a periodic box any term in 
eq.~\eqref{eq:k_budget}, defined as the Fourier transform of the relevant
correlation, can be replaced by the average product of the corresponding Fourier 
coefficients, e.g.  $\Psi(k) = \langle {\hat F}_{R,i}(k) {\hat u}_i^*(k)\rangle$.
In order to get a deeper insight into the origin of the new scaling law,
it is useful to consider the spectrum of the particle back-reaction field 
$F(k) = \langle \hat{F}_{R,i}(k) \hat{F}^*_{R,i}(k) \rangle$.
Figure~\ref{fig:spectral_corr} shows $F(k)$ for the case at $Re_\lambda=55$ and 
$\sigma_R/\eta=0.25$, which is the case with the largest separation between 
Kolmogorov and regularisation scale we have considered. 
In the range of wavenumbers centred at $k\eta \simeq 1$ which are not yet
affected by the regularisation, i.e. $k\sigma_R<1$, $F(k) \simeq {\hat F}^2_0$ is 
roughly constant. This result is somehow expected since the field  
$F_{R,i}(\vx,t)$ is the superposition of Gaussians with variances still 
significantly smaller than the considered scales,  see eq. (\ref{eq:F_regular}).  
The Fourier transform reads
$\displaystyle \hat{F}_{R,i}= -\sum_{p=1}^{N_p} D_{p,i}(t-\varepsilon_R) 
e^{-\frac{1}{2}k^2 \sigma_R^2} \, e^{-\imath k_j x_{p,j}(t-\varepsilon_R)}$
which, apart from the phase, is proportional to $e^{-1/2 k^2 \sigma_R^2}$, hence 
almost constant for $k\sigma_R<1$. The inter-phase momentum coupling $\Psi(k)$ is 
also reported in the figure in comparison with the estimate $\sqrt{F(k) \, E(k)}$ 
(squares). The data show that, where $F(k) \simeq {\hat F}^2_0$, 
$\Psi(k)$ closely matches the curve $\sqrt{F(k) \, E(k)}$. 
It follows that $\Psi(k) \sim \sqrt{F(k) \, E(k)} \sim {\hat F}_0 \sqrt{E(k)}$. 
Then, given the observed $k^{-4}$ scaling for the spectrum, we infer
$\Psi(k) \propto k^{-2}$, as confirmed by the collapse of the data represented by 
circles ($\Psi$), squares ($\sqrt{F(k) \, E(k)}$) and solid line ($k^{-2}$).
In other words, at these scales, the Fourier transform of velocity and backreaction 
are found to be uncorrelated. This suggests that a purely dimensional argument can 
be put forward: neglecting force-velocity correlations in the Fourier modes at 
small scales,  assuming $\Psi(k) \sim {\hat F}_0 \, \hat{u}$,
and introducing the ansatz $\hat{u} \propto k^{\alpha/2}$, the balance of
backreaction $\Psi(k)$ and dissipation 
$D(k) = 2 \nu k^2 E(k) \sim k^2 {\hat u}^\alpha$
leads to the observed scaling law $E(k) \propto k^{-4}$. 

\begin{figure}[b!]
\includegraphics[width=.49\textwidth]{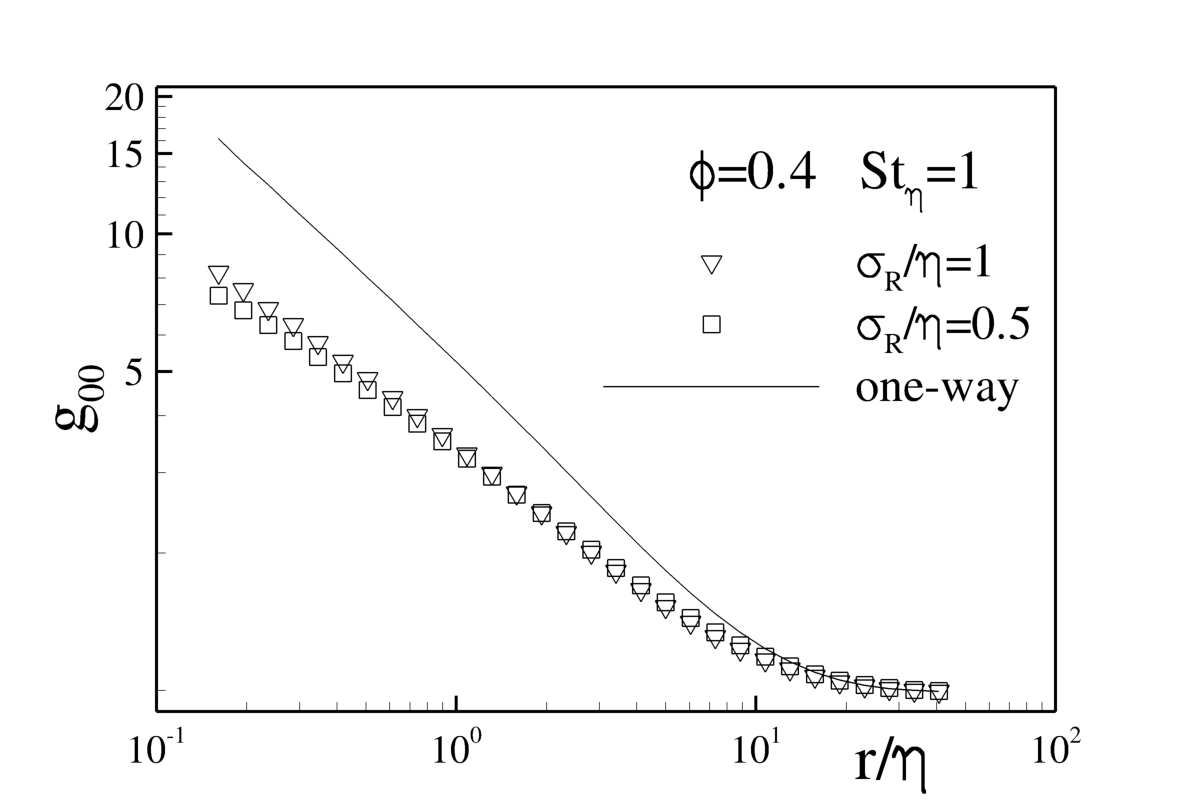}
\includegraphics[width=.49\textwidth]{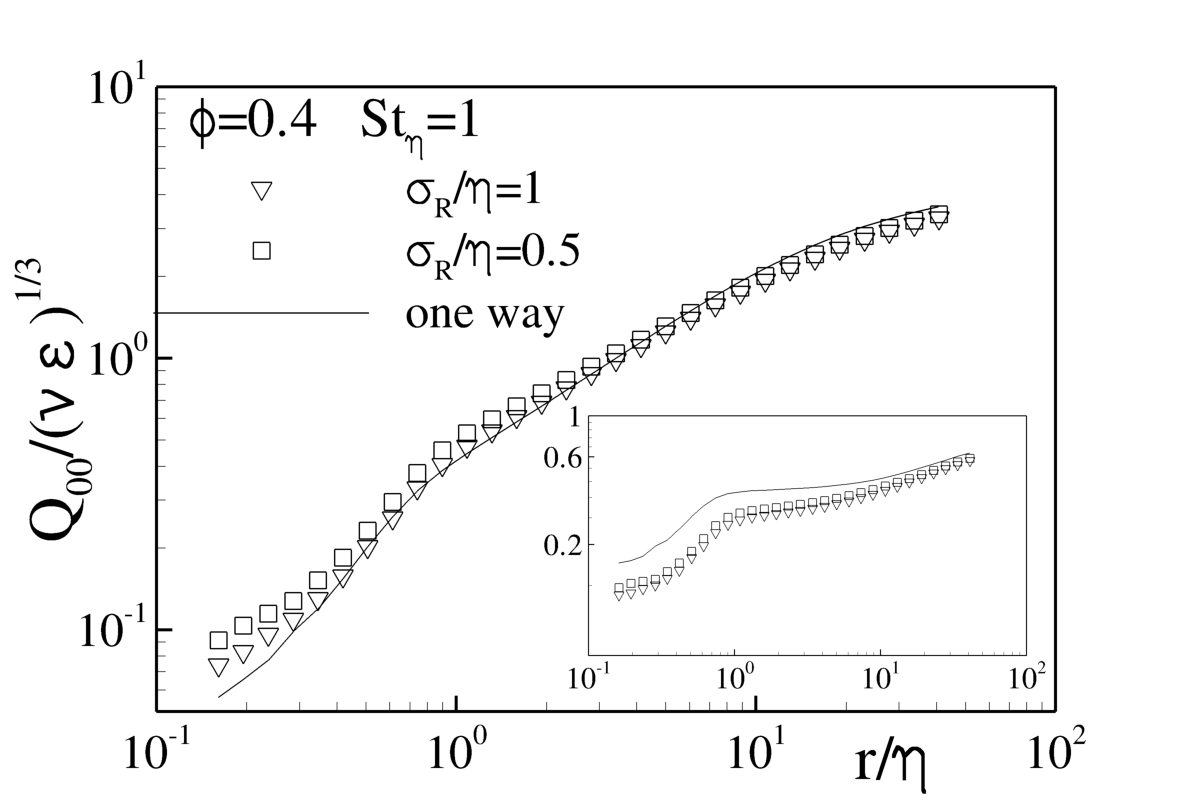}
\caption{Left panel: Radial distribution function $vs.$ separation $r/\eta$.
Data at $Re_\lambda=55$, $St_\eta=1$, $\phi=0.4$: 
$\sigma_R/\eta=1$ ($\bigtriangledown$);
$\sigma_R/\eta=0.5$ ($\square$). For comparison: data in uncoupled conditions
(solid line).
Right panel: Normalised particle pair relative velocity  $vs.$ separation $r/\eta$. 
Inset: product $g_{00}\times Q_{00}$ proportional to the collision rate $vs.$ 
separation.  Data at $Re_\lambda=55$, $St_\eta=1$, $\phi=0.4$:
$\sigma_R/\eta=1$ ($\bigtriangledown$);
$\sigma_R/\eta=0.5$ ($\square$). For comparison: data in uncoupled conditions
(solid line).
\label{fig:clustering}
} 
\end{figure}

From previous studies in the one-way-coupling regime it is well known that
clustering peaks at  $St_\eta ={\cal O}(1)$ \cite{sundaram1997collision,bec2007heavy}. 
Clustering is also observed in the two way coupling regime. It is however 
substantially reduced by the back reaction, as measured by the radial distribution 
function (RDF, see \footnote{The radial distribution function 
$g_{00}(r)$ is the density of particle pairs in a ball ${\cal B}_r$ of radius 
$r$ normalised with the density pairs $n_0=0.5 N_p(N_p-1)/V_0$ in the whole 
domain $V_0$, namely $g_{00}(r) =1/(4\pi r^2 n_0) dN_r/dr$, where $N_r$ is the 
number of pairs in the ball ${\cal B}_r$. The small scales divergence of the 
radial distribution function corresponds to the occurrence of small scale 
clustering. In fact, whenever a scaling law $g_{00}(r) \propto r^{-\alpha}$ with 
positive $\alpha$ occurs, the scaling exponent $\alpha$ measures the 
correlation dimension ${\cal D}_2 = 3 - \alpha$ of the multi-fractal measure 
associated with the particle density \cite{grassberger1983characterization}.})
of the particles shown in figure~\ref{fig:clustering}.
Clustering increases the overall probability that particles 
could collide. Beside clustering, the collision frequency is determined by 
the mean relative velocity of close particles  - a further crucial small scale 
property of the system that needs accurate modelling. Technically, the relevant 
statistical quantity is the average longitudinal velocity difference between two 
particles  
$Q_{00} = \langle \delta v_\parallel(r) | \delta v_\parallel(r) < 0 \rangle$  
where the average is conditioned to negative relative velocity 
$\delta v_\parallel$
\footnote{The collision rate, i.e. the number of collision per unit time and 
volume is given by $\Gamma=2\pi \sigma^2 g_{00}(r=\sigma) Q_{00}(r=\sigma) $ 
where $\sigma = d_p$ is the collision radius, $g_{00}(r=\sigma)$ is the RDF 
evaluated at collision and $Q_{00}(r=\sigma)$ is mean relative velocity of the 
colliding pair, see e.g.~\cite{sundaram1997collision}.}, right panel of
figure~\ref{fig:clustering}. 
The collision probability is  proportional to the product 
$g_{00} \times Q_{00}$, \cite{sundaram1997collision} evaluated
at contact ($r = d_p)$. This object is reported in the inset of the right panel of 
the figure as a function of separation. The present data show that, in the 
relevant range of scales below $\eta$, the two-way coupling may deplete the 
collision frequency since the decrease of the clustering intensity prevails on the 
slight increase of the relative velocity.
%---------------------------------------------------------------------------------
%---------------------------------------------------------------------------------

In conclusion the present Communication highlights  new features of turbulence in 
highly loaded suspensions of tiny, heavy particles. The particles are found to 
drain energy from the carrier flow at the large scales and release it back at 
the small scales. It follows that, in this kind of multiphase flows, 
turbulent fluctuations are unusually forced in the dissipative range. 
The back-reaction stalls the energy cascade and enforces a newly observed 
$E(k)\propto k^{-4}$ scaling law for the energy spectrum at scales order of 
$\eta$, where the particle-injected energy is immediately dissipated by viscosity.
Noteworthy, small scale clustering is depleted by the 
particle-fluid interaction while the relative particle velocity is slightly modified.
Consequently, the collision probability turns out to be reduced. In more general terms, it has been shown that the 
coupling strategy described in the Communication provides a viable technique to 
robustly evaluate small scale statistics in highly loaded particle laden flows. 
The approach, relying on a physical regularisation of the singular force 
feedback, provides convergent result with respect to the regularisation 
parameter allowing a safe evaluation of central observables for heavy loaded 
dilute suspensions. The approach can be easily extended to turbulent flow laden 
with micro-bubbles and to wall bounded flows.

The research received funding from the European Research Council under the 
European Union's Seventh Framework Programme (FP7/2007-2013)/ERC grant agreement 
no. [339446]. Support from  PRACE,  projects FP7 RI-283493 and grant 
no. 2014112647, is acknowledged.

%___________________________________________________________________________

%___________________________________________________________________________
\newpage
\section*{Supplemental Material}
\subsection{The Exact Regularized Point Particle method}
%-------------------------------------------------------------------------------
This section summarizes the physical model used to couple carrier fluid 
and disperse phase for the specific case of periodic boundary conditions considered 
in the simulation of the  homogenous turbulent shear flow.  The reader can refer 
to \cite{Gualtieri_2015} for additional details and more general conditions.

%-------------------------------------------------------------------------------
The carrier fluid fills the domain 
${\cal D}\backslash \Omega$  where $\cal D$ is the periodic flow domain and 
$\Omega(t) = \cup_p \Omega_p(t)$ denotes the region occupied by the 
$N_p$ rigid particles, with $\Omega_p(t)$ the -- small but still finite -- domain 
occupied by the $p$th particle. The motion of the carrier fluid is described by the 
incompressible Navier-Stokes equations with the no-slip condition at 
the particle boundaries and periodic boundary conditions on $\partial {\cal D}$,
%-------------------------------------------------------------------------------
\begin{equation}
\label{eqn:ns_resolved}
\begin{array}{l}
\left.
\begin{array}{l}
\displaystyle \nabla \cdot \vu = 0 \\
\displaystyle \frac{\partial \vu}{\partial t} + \vu \cdot \nabla \vu = 
-\frac {1}{\rho_f} \nabla {\rm p} + \nu \nabla^2 \vu 
\end{array} \right\} \qquad \vx \in {\cal D}\backslash \Omega(t)
\\
~\\
\displaystyle \vu\lvert_{\partial \Omega_p(t)} = \vv_p(\vx)\lvert_{\partial \Omega_p(t)} \qquad \qquad p =1,\ldots,N_p \\
~\\
\displaystyle \vu(\vx,0)=\vu_0(\vx) \qquad \qquad \vx \in {\cal D}\backslash \Omega(0)\ .
\end{array}
\end{equation}
%-------------------------------------------------------------------------------
In equations (\ref{eqn:ns_resolved}), $\vu_0(\vx)$ is the velocity field at time 
$t=0$, $\rho_f$ denotes the fluid density, $\nu$ is the kinematic viscosity, 
and $\vv_p(\vx)$ the velocity of the particle boundary.
In presence of a number of small particles the idea is to relocate the boundary 
condition at the particle surface on a properly defined correction flow field 
for which, in the limit of small particles, an analytical solution can be provided. 
The carrier fluid velocity $\vu$ is decomposed into two parts, 
$\vu(\vx,t)=\vw+\vv$ where the periodic (background) field $\vw(\vx,t)$ is 
assumed to satisfy the equations 
%-------------------------------------------------------------------------------
\begin{equation}
\label{eqn:ns_background}
\begin{array}{l}
\begin{array}{l}
\displaystyle \nabla \cdot \vw = 0 \\
\displaystyle \frac{\partial \vw}{\partial t} + \vF = 
-\frac {1}{\rho_f} \nabla \pi + \nu \nabla^2 \vw \\
\end{array}
\\
\displaystyle \vw(\vx,0)={\vu}_0(\vx)\, ,
\end{array}
\end{equation}
%-------------------------------------------------------------------------------
where $\vx \in {\cal D}$ and
%-------------------------------------------------------------------------------
\begin{equation}
\label{eqn:ns_F}
\vF = \left\{
\begin{array}{ll}
\vu \cdot \nabla  \vu & \qquad \mbox{for} \,\, \vx \in  
{\cal D}\backslash\Omega(t)\\ \\
\vv_p \cdot \nabla \vv_p & \qquad \mbox{for} \,\, \vx \in  \Omega(t)
\end{array}
\right.
\end{equation}
%-------------------------------------------------------------------------------
is a field reproducing the convective term of the Navier-Stokes equation 
in ${\cal D}\backslash \Omega$. For the time being, in (\ref{eqn:ns_background}) the 
convective term is considered as prescribed 
and the no-slip condition at the particle surface has been removed.
In fact, the no-slip  boundary condition at the 
particle surface is recovered when considering the particle perturbation field 
$\vv(\vx,t)$ which satisfies the \emph{linear} unsteady Stokes 
problem (the complete non-linear term has been retained in the equation for $\vw$) 
%-------------------------------------------------------------------------------
\begin{equation}
\label{eqn:unsteady_stokes}
\begin{array}{l}
\left.
\begin{array}{l}
\displaystyle \nabla \cdot \vv = 0 \\
\displaystyle \frac{\partial \vv}{\partial t}
=-\frac{1}{\rho_f} \nabla {\rm q} +\nu \nabla^{2}{\vv}  
\end{array}
\right\} \qquad \vx \in {\cal D}\backslash \Omega(t)
\\
~\\
\displaystyle \vv\lvert_{\partial \Omega_p(t)}=\vv_p(\vx)\lvert_{\partial \Omega_p(t)} - 
\vw\lvert_{\partial \Omega_p(t)} \qquad p = 1, \ldots N_p \\
~\\
\displaystyle \vv(\vx,0)=0 \qquad \vx \in  {\cal D}\backslash \Omega(0)\, .
\end{array}
\end{equation}
%-------------------------------------------------------------------------------
The boundary integral representation of the solution to the unsteady Stokes equations (\ref{eqn:unsteady_stokes})
can be expressed in terms of multipoles. In the limit of small particles  the far field reduces to
%-------------------------------------------------------------------------------
\begin{equation}
\label{eqn:unsteady_stokes_far_field}
v_i(\vx,t)=-\sum_p \int_0^t D^p_j(\tau) G_{ij}(\vx,\vx_p,t,\tau) \, d\tau \, ,
\end{equation}
%-------------------------------------------------------------------------------
where $G_{ij}(\vx,\vxi,t,\tau)$ is the unsteady Stokeslet, i.e. 
the fluid velocity ($i$th direction) at position  $\vx$ and time $t$ 
due to the singular forcing $\delta(\vx-\vxi)\delta(t-\tau)$  ($j$th direction) 
applied at point $\vxi$ and at time $\tau$ and  $D^p_j(\tau)$ are the Cartesian components
of the hydrodynamic force on the particle.  The partial differential equation whose solution is 
given by (\ref{eqn:unsteady_stokes_far_field}) reads
%-------------------------------------------------------------------------------
\begin{equation}
\label{eqn:unsteady_stokes_singular}
\begin{array}{l}

\displaystyle \frac {\partial \vv}{\partial t} - \nu \nabla^2 {\bf v} + 
\frac{1}{\rho_f} \nabla {\rm q} =  - \frac{1}{\rho_f} \sum_p \vD_p(t)
\, \delta\left[ {\bf x} -{\bf x}_p(t) \right]  \\ \\

\displaystyle \vv(\vx,0)=0 \, .
\end{array}
\end{equation}
%-------------------------------------------------------------------------------
A regularized solution of equation (\ref{eqn:unsteady_stokes_singular})
can be achieved by reasoning in terms of the  vorticity field 
$\vzeta = \nabla \times \vv$ which obeys a (vector) diffusion equation. 
The solution is
%-------------------------------------------------------------------------------
\begin{equation}
\label{eqn:sol_vort_forced}
\vzeta(\vx,t) = \frac{1}{\rho_f}\int_0^{t} 
\vD_p(\tau) \times \nabla g\left[ \vx-\vx_p(\tau),t-\tau\right] d\tau \, ,
\end{equation}
%-------------------------------------------------------------------------------
%-------------------------------------------------------------------------------
where  $g(\vx-\vxi,t-\tau)$ is a Gaussian function with time dependent 
variance $\sigma(t-\tau) = \sqrt{2 \nu (t-\tau)}$.  The (still) singular field $\vzeta$ 
is regularized using  a temporal cut-off  $\epsilon_R$ leading to its splitting into a regular and a 
singular component  $\vzeta(\vx,t) = \vzeta_R(\vx,t;\epsilon_R) + \vzeta_S(\vx,t;\epsilon_R)$
%-------------------------------------------------------------------------------
where 
%-------------------------------------------------------------------------------
\begin{equation}
\label{eqn:vort_regular}
\vzeta_R(\vx,t) = \frac{1}{\rho_f}
\int_0^{t - \epsilon_R}
\vD_p(\tau) \times \nabla g\left[ \vx-\vx_p(\tau),t-\tau\right] d\tau 
\end{equation}
%-------------------------------------------------------------------------------
is  smooth with smallest spatial scale given by
$\sigma_R=\sigma(\epsilon_R) = \sqrt{2 \nu \epsilon_R}$. 
It obeys a forced diffusion 
equation where the forcing is applied at the slightly earlier time $t-\epsilon_R$,
%-------------------------------------------------------------------------------
\begin{eqnarray}
\label{eqn:pde_vort_regular}
\frac{\partial \vzeta_R}{\partial t} & - & \nu  \nabla^2 \vzeta_R = \nonumber \\
& - & \frac{1}{\rho_f} \nabla \times \vD_p(t-\epsilon_R)
g\left[\vx -\vx_p(t-\epsilon_R),\epsilon_R 
\right]  
\end{eqnarray}
%-------------------------------------------------------------------------------
with $\vzeta_R(\vx,0)=0$. The associated velocity field obeys the forced unsteady 
Stokes equation
%-------------------------------------------------------------------------------
\begin{eqnarray}
\label{eqn:velo_reg_eq}
\frac{\partial \vv_R}{\partial t} & - &  \nu \nabla^2 \vv_R 
+\frac{1}{\rho_f}\nabla {\rm q}_R  =  \nonumber \\
& - & \frac{1}{\rho_f} \vD_p(t-\epsilon_R) \,
g\left[ \vx-\vx_p(t-\epsilon_R),\epsilon_R \right]
\end{eqnarray}
%-------------------------------------------------------------------------------
for the solenoidal field $\vv_R$
that can be split in terms of a pseudo-velocity,
%-------------------------------------------------------------------------------
\begin{eqnarray}
\label{eqn:pseuso_velo_reg_eq}
\frac{\partial \vv_{\vzeta_{R}}}{\partial t} & - & \nu \nabla^2 \vv_{\vzeta_{R}} 
= \nonumber \\ & - & \frac{1}{\rho_f}
\vD_p(t-\epsilon_R) \, g\left[ \vx-\vx_p(t-\epsilon_R),\epsilon_R \right]
\end{eqnarray}
%-------------------------------------------------------------------------------
governed be the unsteady diffusion operator plus a gradient correction required to enforce 
solenoidality.  The highly localized singular contribution $\vv_S$, which  cannot be represented 
on a discrete grid,  is successively reintroduced in the  field as soon as it diffuses sufficiently to 
reach the smallest physically relevant scales.  The regularized (solenoidal) fluid velocity in presence of the 
particles is $\vu_R=\vw+\vv_R$ and is governed by the equation
%-------------------------------------------------------------------------------
\begin{eqnarray}
\label{eqn:ns_regularized_filtered}
\frac{\partial \vu}{\partial t} & + & \vu \cdot \nabla \vu =  
 - \frac {1}{\rho_f} \nabla p + \nu \nabla^2 \vu \nonumber \\
& - & \frac{1}{\rho_f} \sum_p^{N_p}
\vD_p(t-\epsilon_R) \, g\left[ \vx-\vx_p(t-\epsilon_R),\epsilon_R \right]  \ .
\end{eqnarray}
%-------------------------------------------------------------------------------
%-------------------------------------------------------------------------------
%-------------------------------------------------------------------------------
%-------------------------------------------------------------------------------
\subsection{The Removal of Particle Self-interaction in the Evaluation of the 
Hydrodynamic Force}
%-------------------------------------------------------------------------------
The hydrodynamic force acting on a small particle of diameter $d_p$ and density 
$\rho_p\gg\rho_f$ reduces to the Stokes drag~\cite{gatignol1983faxen,Maxey_1983}, 
%-------------------------------------------------------------------------------
\begin{equation}
\vD_p(t) =   6\pi\mu a_p \left[\tilde \vu(\vx_p,t) -\vv_p(t) \right] 
\end{equation}
%-------------------------------------------------------------------------------
The velocity $\tilde \vu(\vx_p,t)$ is the fluid velocity, 
at the particle position, in absence of the particle self-disturbance, i.e. 
$\tilde \vu_p$ must account for the background turbulent flow altered by the disturbances
generated by all the other particles except the $p$th one.
In the two-way coupling regime where the particle back-reaction 
modifies the carrier flow, the calculation of $\tilde \vu_p$ needs
the removal from the field $\vu(\vx,t)$ of the particle self-interaction contribution.
In the ERPP approach the (regularised) disturbance flow induced at time $t$
and position $\vx$ by a particle located at $\vx_0$, i.e. $\vv_R(\vx-\vx_0,t)$,
is known in closed form. 
The actual hydrodynamic force on the $p$th particle can be evaluating by 
subtracting from $\vu(\vx_p,t)$ the value 
$\vv_R[\vx_p(t)- \vx_p(t-\Delta t), \Delta t]$ induced at time $t$ 
at the current particle position $\vx_p(t)$ by the same particle when
it was placed at $\vx_p(t-\Delta t)$,
%-------------------------------------------------------------------------------
\begin{eqnarray}
\label{eqn:self_dist}
\vv_R(\vx,t_{n+1}) & = & \frac{1}{\left(2 \pi \sigma^2\right)^{3/2}} 
\left\{\left[e^{-\eta^2} -\frac{f(\eta)}{2\eta^3}\right]  \vD^n \right. \nonumber \\
& - & \left. \left(\vD^n \cdot \hat{\vvr}\right)
\left[e^{-\eta^2} -\frac{3f(\eta)}{2\eta^3}\right] \hat{\vvr} \right\} \, ,
\end{eqnarray}
%-------------------------------------------------------------------------------
where $\vD^n=\vD(t_n-\epsilon_R)$,
$\vvr=\vx-\vx_p(t_n-\epsilon_R)$, the hat denotes the unit vector $\hat{\vvr}=\vvr/r$,
$\eta=r/\sqrt{2} \sigma$ is the dimensionless distance with
$\sigma=\sqrt{2\nu(\epsilon_R + \Delta t)}$  and
$\displaystyle f(\eta)=\frac{\sqrt{\pi}}{2}\mbox{erf}(\eta)-\eta e^{-\eta^2}$,
see~\cite{Gualtieri_2015} for the formal derivation of equation \eqref{eqn:self_dist}.
%-------------------------------------------------------------------------------
This procedure can be straightforwardly extended to the Runge-Kutta algorithm 
employed in the present Letter to integrate in time the equations of the 
carrier and of the disperse phase.
%-------------------------------------------------------------------------------
%-------------------------------------------------------------------------------
%-------------------------------------------------------------------------------
%-------------------------------------------------------------------------------
\subsection{The homogeneous shear flow}
%-------------------------------------------------------------------------------
%-------------------------------------------------------------------------------
The homogeneous shear flow consists in a turbulent flow into a periodic box
in which velocity fluctuations are fed by an imposed mean velocity profile,
see~\cite{pumir1996turbulence,gualtieri2002scaling} for more details.
The velocity field $\vv$ is decomposed into a mean flow $\vU=S x_2 \, \ve_1$ and a 
fluctuation $\vu$ where $\ve_1$ is the unit vector in the $x_1$ direction 
(streamwise), $x_2$ denotes the coordinate in the direction of the mean constant 
shear $S$ (transverse direction) and $x_3$ is the spanwise direction. 
The dynamics of the velocity fluctuations in a deforming coordinate 
system convected by the mean flow according to the transformation of variables 
$\xi_1=x_1 - S t x_2; \quad \xi_2=x_2; \quad \xi_3=x_3; \quad \tau=t$ is 
described by the incompressible Navier-Stokes equations~\cite{rogallo1981numerical}
%----------------------------------------------------------------------------------
\begin{equation}
\frac{\partial \vu}{\partial \tau}+\vu \cdot \nabla \vu =
-\frac{1}{\rho}\nabla p +\nu \nabla^{2}\vu -S u_2 \ve_1 \, .
\end{equation}
%----------------------------------------------------------------------------------
In the homogeneous shear flow the Reynolds shear stresses 
$\langle u_1 \, u_2 \rangle$ extract energy from the mean flow and feeds 
the energy cascade up to viscous dissipation according to the balance 
$-S \langle u_1 \, u_2 \rangle = \epsilon$ where $\epsilon$ is energy 
dissipation rate. Turbulent fluctuations are spatially homogeneous and 
statistically stationary in time. Beyond the integral scale 
$L_0 = \left( 2 K \right)^{3/2}/\epsilon$ and Kolmogorov dissipative scale 
$\eta=\left( \nu^3/\epsilon\right)^{1/4}$, the homogeneous shear flow features the 
so-called shear scale $L_S = \sqrt{\epsilon/S^3}$, where $K$ is the average 
turbulent kinetic energy. Due to the shear, turbulent fluctuations are strongly 
anisotropic at large scales $L_S \ll r \ll L_0$ (production range) where the production
associated with the Reynolds stresses overwhelms the other mechanisms.
%they are driven by the production mechanisms associated to the Reynolds stresses. 
At small scales, $\eta \ll r \ll L_S$ (isotropy recovery range) inertial 
energy transfer prevails leading to the classical Kolmogorov energy cascade. 
Two dimensionless parameters characterise the flow, 
the Corrsin parameter, $S_c = \sqrt{S^2 \nu/\epsilon} = \left(\eta/L_S \right)^{2/3}$,
and the shear strength, $S^* = 2 K  S/\epsilon = \left( L_0/L_S \right)^{2/3}$, 
which ratio corresponds to the classical Taylor-Reynolds number, 
$Re_\lambda = 2 K /\sqrt{\nu \epsilon}=S^*/S_c$.

%%___________________________________________________________________________

\begin{thebibliography}{40}

\bibitem{Johansen_2007}
{\sc Johansen,  Oishi,  Mac Low,  Klahr, Henning, Youdin,},
"Rapid planetesimal formation in turbulent circumstellar disks",
Nature, {\bf 448}, 2007.

\bibitem{Takeuchi_2002}
{\sc Takeuchi, Lin },
"Radial flow of dust particles in accretion disks",
The Astrophysical Journal, {\bf 581}, 2002.

\bibitem{mitra2013can}
{\sc Mitra, Wettlaufer, Brandenburg},
"Can planetesimals form by collisional fusion?",
The Astrophysical Journal, {\bf 773}, 2013.

\bibitem{pan2014turbulence}
{\sc Pan, Padoan},
"Turbulence-induced Relative Velocity of Dust Particles. IV. The Collision Kernel", The Astrophysical Journal, {\bf 797}, 2014.

\bibitem{fu2014effects}
{\sc Fu, Li, Lubow, Li, Liang},
"Effects of dust feedback on vortices in protoplanetary disks",
The Astrophysical Journal Letters, {\bf 795}, 2014.

\bibitem{lanotte2009cloud}
{\sc Lanotte, Seminara, Toschi},
"Cloud droplet growth by condensation in homogeneous isotropic turbulence",
Journal of the Atmospheric Sciences, {\bf 66}, 2009.

\bibitem{Falkovich_2002}
{\sc Falkovich, Fouxon, Stepanov},
"Acceleration of rain initiation by cloud turbulence",
Nature, {\bf 419}, 2002.

\bibitem{Shaw_2003}
{\sc Shaw},
"Particle-turbulence interactions in atmospheric clouds",
Annual Review of Fluid Mechanics, {\bf 35}, 2003.

\bibitem{Marmottant_2004}
{\sc Marmottant, Villermaux},
"On spray formation", Journal of fluid mechanics, {\bf 498}, 2004.

\bibitem{lin_1998}
{\sc Lin, Reitz},
"Drop and spray formation from a liquid jet",
Annual Review of Fluid Mechanics, {\bf 30}, 1998.

\bibitem{jenny2012modeling}
{\sc Jenny, Roekaerts, Beishuizen},
"Modeling of turbulent dilute spray combustion",
Progress in Energy and Combustion Science, {\bf 38}, 2012.

\bibitem{Abraham_1998}
{\sc Abraham},
"The generation of plankton patchiness by turbulent stirring",
Nature, {\bf 391}, 1998.

\bibitem{dunkel2013fluid}
{\sc Dunkel, Heidenreich, Drescher, Wensink, B{\"a}r, Goldstein},
"Fluid dynamics of bacterial turbulence",
Physical review letters, {\bf 110}, 2013.

\bibitem{lance1991turbulence}
{\sc Lance, Bataille},
"Turbulence in the liquid phase of a uniform bubbly air--water flow",
Journal of Fluid Mechanics, {\bf 222}, 1991.

\bibitem{ferguson2015internal}
{\sc Ferguson, Kirkpatrick},
"Internal combustion engines: applied thermosciences",
John Wiley \& Sons, 2015.

\bibitem{elghobashi1994predicting}
{\sc Elghobashi},
"On predicting particle-laden turbulent flows",
Applied Scientific Research, {\bf 52}, 1994.

\bibitem{elgo_map}
{\sc Elgobashi},
"An updated classification map of particle-laden turbulent flows",
IUTAM Symposium on Computational Approaches to Multiphase Flow, 2006.

\bibitem{Balachandar_2010}
{\sc Balachandar, Eaton},
"Turbulent dispersed multiphase flow",
Annual Review of Fluid Mechanics, {\bf 42}, 2010.

\bibitem{bec2007heavy}
{\sc Bec, Biferale, Cencini, Lanotte, Musacchio, Toschi},
"Heavy particle concentration in turbulence at dissipative and inertial scales",
Physical review letters, {\bf 98}, 2007.

\bibitem{saw2008inertial}
{\sc Saw, Shaw, Ayyalasomayajula, Chuang, Gylfason},
"Inertial clustering of particles in high-Reynolds-number turbulence",
Physical review letters, {\bf 100}, 2008.

\bibitem{young1997theory}
{\sc Young, Leeming},
"A theory of particle deposition in turbulent pipe flow",
Journal of Fluid Mechanics, {\bf 340}, 1997.

\bibitem{fessler1994preferential}
{\sc Fessler, Kulick, Eaton},
"Preferential concentration of heavy particles in a turbulent channel flow",
Physics of Fluids, {\bf 6}, 1994.

\bibitem{chun2005clustering}
{\sc Chun, Koch, Rani, Ahluwalia, Collins},
"Clustering of aerosol particles in isotropic turbulence",
Journal of Fluid Mechanics, {\bf 536}, 2005.

\bibitem{Gualtieri_2009}
{\sc Gualtieri, Picano, Casciola},
"Anisotropic clustering of inertial particles in homogeneous shear flow",
Journal of Fluid Mechanics, {\bf 629}, 2009.


\bibitem{Gualtieri_2009}
{\sc Gualtieri, Picano, Casciola},
"Anisotropic clustering of inertial particles in homogeneous shear flow",
Journal of Fluid Mechanics, {\bf 629}, 2009.

\bibitem{picano2010anomalous}
{\sc Picano, Sardina, Gualtieri, Casciola},
"Anomalous memory effects on transport of inertial particles in turbulent jets",
Physics of Fluids, {\bf 22}, 2010.

\bibitem{kolmogorov1941local}
{\sc Kolmogorov},
"The local structure of turbulence in incompressible viscous fluid for very large Reynolds numbers",
Dokl. Akad. Nauk SSSR, {\bf 30}, 1941.

\bibitem{frisch1995turbulence}
{\sc Frisch},
"Turbulence: the legacy of AN Kolmogorov",
Cambridge university press, 1995.

\bibitem{Gualtieri_2015}
{\sc Gualtieri, Picano, Sardina, Casciola},
"Exact regularized point particle method for multiphase flows in the two-way coupling regime", Journal of Fluid Mechanics, {\bf 773}, 2015.

\bibitem{SI}
{\sc Gualtieri, Battista, Casciola},
Supplementary Information.

\bibitem{crowe1977particle}
{\sc Crowe, Sharma, Stock},
"The particle-source-in cell (PSI-CELL) model for gas-droplet flows",
Journal of Fluids Engineering, {\bf 99}, 1977.

\bibitem{Gualtieri_2013}
{\sc Gualtieri, Picano, Sardina, Casciola},
"Clustering and turbulence modulation in particle-laden shear flows",
Journal of Fluid Mechanics, {\bf 715}, 2013.

\bibitem{boivin1998direct}
{\sc Boivin, Simonin, Squires},
"Direct numerical simulation of turbulence modulation by particles in isotropic turbulence",
Journal of Fluid Mechanics, {\bf 375}, 1998.

\bibitem{gatignol1983faxen}
{\sc Gatignol},
"The Fax{\'e}n formulas for a rigid particle in an unsteady non-uniform Stokes-flow",
Journal de M{\'e}canique th{\'e}orique et appliqu{\'e}e, {\bf 2}, 1983.

\bibitem{Maxey_1983}
{\sc Maxey, Riley},
"Equation of motion for a small rigid sphere in a nonuniform flow",
Physics of Fluids, {\bf 26}, 1983.

\bibitem{casciola2003scale}
{\sc Casciola, Gualtieri, Benzi, Piva},
"Scale-by-scale budget and similarity laws for shear turbulence",
Journal of Fluid Mechanics, {\bf 476}, 2003.

\bibitem{eyink2006onsager}
{\sc Eyink, Sreenivasan},
"Onsager and the theory of hydrodynamic turbulence",
Reviews of modern physics, {\bf 78}, 2006.

\bibitem{sundaram1997collision}
{\sc Sundaram, Collins},
"Collision statistics in an isotropic particle-laden turbulent suspension. Part 1. Direct numerical simulations",
Journal of Fluid Mechanics, {\bf 335}, 1997.

\bibitem{grassberger1983characterization}
{\sc Grassberger, Procaccia},
"Characterization of strange attractors",
Physical review letters, {\bf 50}, 1983.

\bibitem{pumir1996turbulence}
{\sc Pumir},
"Turbulence in homogeneous shear flows", Physics of Fluids, {\bf 8}, 1996.

\bibitem{gualtieri2002scaling}
{\sc Gualtieri, Casciola, Benzi, Amati, Piva},
"Scaling laws and intermittency in homogeneous shear flow,
Physics of Fluids, {\bf 14}, 2002.

\bibitem{rogallo1981numerical}
{\sc Rogallo},
"Numerical experiments in homogeneous turbulence", NASA TM-81315, 1981.

\end{thebibliography}
\end{document}